# Effects of heat dissipation on unipolar resistance switching in Pt/NiO/Pt capacitors

S. H. Chang, S. C. Chae, S. B. Lee, C. Liu, and T. W. Noh[*]
*ReCOE & FPRD, Department of Physics and Astronomy, Seoul National University, Seoul 151-747, Korea*

J. S. Lee and B. Kahng
*Department of Physics and Astronomy, Seoul National University, Seoul 151-747, Korea*

J. H. Jang and M. Y. Kim
*School of Materials Science and Engineering, Seoul National University, Seoul 151-744, Korea*

D.-W. Kim
*Department of Applied Physics, Hanyang University, Ansan, Gyeonggi 426-791, Korea*

C. U. Jung
*Department of Physics, Hankuk University of Foreign Studies, Yongin, Gyeonggi 449-791, Korea*

We fabricated Pt/NiO/Pt capacitor structures with various bottom electrode thicknesses, $t_{BE}$, and investigated their resistance switching behaviors. The capacitors with $t_{BE} \geq 50$ nm exhibited typical unipolar resistance memory switching, while those with $t_{BE} \leq 30$ nm showed threshold switching. This interesting phenomenon can be explained in terms of the temperature-dependent stability of conducting filaments. In particular, the thinner $t_{BE}$ makes dissipation of Joule heat less efficient, so the filaments will be at a higher temperature and become less stable. This study demonstrates the importance of heat dissipation in resistance random access memory.

[*] Corresponding Author. E-mail: twnoh@snu.ac.kr

Resistance switching phenomena, observed in numerous materials,[1,2] have regained a great deal of attention recently due to their potential application in nonvolatile memory devices called resistance random access memory (RRAM).[3–9] Achieving good scalability is an important issue for meeting the current demands for device miniaturization.[10–12] One of the important scalability issues is reducing the bottom electrode thickness, $t_{BE}$, to make the etching process easier.[13] Reducing $t_{BE}$ also lowers the device fabrication cost, especially when using expensive electrode materials such as Pt or other noble metals.

Many binary transition oxides such as NiO, TiO$_2$, and Fe$_2$O$_3$ exhibit unipolar resistance switching.[5–8] Although the mechanism is still somewhat ambiguous, it is widely accepted that unipolar resistance switching is due to the formation and rupture of conducting filamentary paths under external bias.[1,6,14–19] It is also generally assumed that the rupturing process of the conducting filaments may be closely related to Joule heating.[17–19] If this is true, we could control Joule heating effects in unipolar resistance switching by changing the thermal properties of the RRAM device structure, which is typically made with metal/oxide/metal. In particular, $t_{BE}$ could significantly affect the thermal heat dissipation process through the bottom electrode.

In this letter, we investigated resistance switching behaviors of Pt/NiO/Pt capacitor structures as a function of $t_{BE}$. We found that capacitors with $t_{BE} \leq 30$ nm exhibited volatile resistance switching behaviors called threshold switching. On the other hand, all capacitors with $t_{BE} \geq 50$ nm exhibited typical unipolar memory switching behaviors. We explained this interesting $t_{BE}$ dependence in terms of the thermal stability of conducting filamentary paths, which are closely related to heat dissipation through the bottom electrode. This result indicates that thermal heat dissipation through the electrodes is crucial for RRAM, just as it is for phase change random access memory.[20]

We fabricated Pt/NiO/Pt capacitor structures by growing polycrystalline NiO films on Pt/TiO$_x$/SiO$_2$/Si substrates. We used e-beam evaporation to deposit the Pt bottom electrode layers with a $t_{BE}$ of 10 – 50 nm. The thinnest Pt layer with $t_{BE}$ = 10 nm had a metallic behavior with electrical conductivity of 2.16×10$^6$ Ω$^{-1}$m$^{-1}$ at 300 K, similar to the reported value of a very thin Pt film.[21] For thicker Pt bottom electrodes with a $t_{BE}$ of 50 – 200 nm, we used commercially available platinized silicon substrates (Inostek Inc., Seoul, Korea) that were grown by sputtering. We deposited Ni films on these substrates using e-beam evaporation, and oxidized them into NiO layers using thermal oxidation at 450 °C at ambient air pressure for 1 hr.[22] The NiO layers were approximately 60 nm thick. For top electrodes, we deposited Au and Pt layers (30 and 10 nm thick, respectively) using e-beam evaporation with a shadow mask. All of the Pt/NiO/Pt capacitors used in this study had an electrode area of 46 × 46 μm.

We investigated the resistance switching characteristics of our Pt/NiO/Pt capacitors by measuring the current-voltage (*I-V*) curves with a conventional two-probe measurement system (Agilent 4155C Semiconductor Parameter Analyzer, Agilent Technologies, Santa Clara, CA). To avoid a complete dielectric breakdown, we used compliance current to set a limit to the current flow.

Interestingly, we observed two types of unipolar resistance switching behaviors in our Pt/NiO/Pt capacitors: memory and threshold switching. As shown in Fig. 1(a), a capacitor with $t_{BE}$ = 150 nm has an *I-V* curve typical of resistance memory switching. When we increased *V*, the resistance changed to a low-resistance state (LRS) at the setting voltage, marked with "1" in the figure. The capacitor was then in a LRS. When we increased *V* from zero again, it changed to a high-resistance state (HRS) at the resetting voltage, marked with "2". On the other hand, a capacitor with $t_{BE}$ = 20 nm had an *I-V* curve of typical threshold switching, as shown in Fig 1(b).[5,23] Although this capacitor exhibited an interesting resistance change from the HRS to the LRS, it could be maintained in the LRS. Therefore, it was not suitable for use as a nonvolatile memory device since it did not have bistable states without an external voltage. Although not shown here, both samples had symmetric *I-V* curves along the bias polarity, i.e., both resistance switching characteristics were unipolar.

We found that the resistance switching behavior varied systematically depending on the value of $t_{BE}$. As shown in Fig. 1(c), all of the capacitors with $t_{BE} \leq 30$ nm exhibited threshold switching behaviors while all of the capacitors with $t_{BE} \geq 50$ nm exhibited resistance memory switching behaviors. This indicates that $t_{BE}$ is a key parameter for controlling the type of resistance switching.

To clarify the origin of the $t_{BE}$-dependent change of the type of resistance switching, we considered three possibilities: the chemical stoichiometry of the NiO films, the structural and chemical property changes near the bottom electrode interface, and the heat dissipation effects. While most NiO films have resistance memory switching characteristics,[1,5,15,19,22,23] there have been several reports of threshold switching in relative Ni-deficient NiO films.[5,23] However, in our case, we deposited Ni films and oxide onto NiO layers under the same conditions. Therefore, a variation in chemical stoichiometry is not a plausible candidate to explain the change in the type of resistance switching illustrated in Fig. 1.

We also investigated cross-sectional images and depth profiles of our Pt/NiO/Pt capacitors using a transmission electron microscope (TEM) equipped for energy-dispersive X-ray spectroscopy, although we do not show the results here. We found that the structural properties near the interface between the bottom Pt electrode and the NiO layer do not vary significantly in terms of $t_{BE}$. In addition, solid-state diffusions of Ni, Ti, O, Pt, and Si ions were found to be rather small. If such diffusion were significant, chemical compositions of the top and bottom electrode interfaces would be different, and their junction-related parameters such as Schottky barrier height would vary.[24] This would mean that the *I-V* characteristics would be asymmetric,[24] contrary to our observations. Therefore, the structural and chemical properties near the bottom electrode interface do not play an important role.

Joule-heating effects are widely considered to be important for the resetting process, i.e., the change from LRS to HRS. There is no direct evidence of such effects, but some research has presented an argument based on the time scale of the resetting process. A typical reported value of the characteristic time for the resetting process is a few microseconds for capacitors that are sub-micron or larger in size. This is much longer than the characteristic time for the change from a HRS to LRS, which typically about 10 ns.[12,18] A relatively long time scale is consistent with a slow thermal process. With a thinner $t_{BE}$, thermal dissipation though the bottom electrode will decrease so that conducting filaments inside the NiO capacitor can reach a higher temperature and rupture more easily. Therefore, if these Joule-heating effects are significant, they could affect the thermal stability of conducting filamentary channels.

To confirm the $t_{BE}$-dependent thermal dissipation effects, we calculated a three-dimensional heat flow by solving the following Fourier equation using finite element analysis:[25,26]

$$\rho C_P \frac{\partial T}{\partial t} = k \nabla^2 T + Q \tag{1}$$

where $\rho$, $C_p$, and $k$ are the density (g/cm$^3$), specific heat (J/g K), and thermal conductivity (W/cm K), respectively, of the constituents. The parameter $Q$ is the heat flux density (W/cm$^3$) supplied by the electric current through the conducting filaments. Table I contains values of the constituents in our NiO capacitors.[27,28] Recently, there have been several reports on the formation of conducting channels studied with scanning probe microscopes.[6,14] The reported size of a conducting channel is typically in the range of 5 – 70 nm.

Figure 2(a) shows a schematic diagram that we used in our calculations. For the sake of simplicity, we assumed that only one rectangular filament, 30 × 30 × 60 nm, is formed inside the capacitor. The top and bottom of the capacitor are in contact with the air and Si substrates, respectively, and all components are assumed to be at a room temperature of 300 K. The resistance of the filament is 30 – 40 μΩ·cm, and it is assumed to be under an external bias of 2 V. In our simulations, we used actual values for the dimensions of the Au/Pt top electrode, NiO, $TiO_2$ adhesion, and $SiO_2$ layers. We calculated two cases with $t_{BE}$ values of 10 and 200 nm.

As shown in Fig. 2(b), the filament's temperature becomes saturated at a time scale of about 1 μs for both cases, which agrees with earlier experimental work.[12,18] Figures 2(c) and (d) show contour plots of the temperature gradients in selected cross-sectional areas near the conducting filament for a scale of 320 × 330 nm. The filament temperature for $t_{BE}$ = 10 nm is about 910 K, which is approximately 200 K higher than that for $t_{BE}$ = 200 nm. This result indicates that the conducting filament could be less thermally stable in a capacitor with a thinner $t_{BE}$.

Note that the thermal conductivity of $SiO_2$ (0.014 W/cm K) is about 50 times less than that of Pt (0.72 W/cm K). Therefore, the heat dissipation through the thinner bottom electrode should be much less efficient. Figures 3(a) and (b) show vector representations of heat flow, *i.e.* -k∇T, for the cases of $t_{BE}$ = 200 and $t_{BE}$ = 10 nm, respectively. In Fig. 3(a), the heat flow vectors inside the thicker bottom electrode are pointing in nearly every direction, and remain significant even for the region about 100 nm below the NiO/Pt interface. On the other hand, in Fig. 3(b), the heat flow vectors inside the thinner bottom electrode are confined mostly to a very narrow region near the bottom electrode, indicating that heat dissipation occurs though the thin bottom electrode and is very inefficient. These simulation results suggest the importance of parameters such as $t_{BE}$ that control the heat dissipation process in real RRAM devices.

In summary, we found that the bottom electrode thickness of Pt/NiO/Pt capacitors plays an important role in their resistance switching behaviors. Resistance memory switching phenomena become unstable and turn into threshold switching when there are thinner bottom electrodes. We explained these phenomena in terms of the thermal stability of the conducting filaments. This work demonstrates the importance of controlling heat dissipation in unipolar resistance memory switching and associated devices.

This work was supported financially by the Creative Research Initiatives (Functionally Integrated Oxide Heterostructure) of the Korean Science and Engineering Foundation (KOSEF).

TABLE I. Parameter values of the Fourier heat equation for Pt, Au, Ni, NiO, $SiO_2$, and $TiO_2$.

| Material | Density $\rho$ (g/cm$^3$) | Specific heat $C_p$ (J/g K) | Thermal conductivity $k$ (W/cm K) |
|---|---|---|---|
| Pt | 22 | 0.13 | 0.72 |
| Au | 19 | 0.13 | 3.2 |
| Ni | 8.9 | 0.44 | 0.91 |
| NiO | 6.7 | 0.59 | 0.35 |
| $SiO_2$ | 2.2 | 0.74 | 0.014 |
| $TiO_2$ | 4.2 | 0.69 | 0.13 |

**Figure captions**

Fig. 1. Resistance switching behaviors of Pt/NiO(60 nm)/Pt capacitors with an area of 46 × 46 μm. (a) A capacitor with $t_{BE}$ = 150 nm shows typical unipolar resistance memory switching behavior, while (b) a capacitor with $t_{BE}$ = 20 nm shows threshold switching behavior. (c) Types of resistance switching behaviors in terms of the Pt bottom electrode thickness. All the capacitors with $t_{BE} \leq 30$ nm showed threshold switching, while all the capacitors with $t_{BE} \geq 50$ nm showed unipolar resistance memory switching.

Fig. 2 (color online). Results of a finite element analysis for thermal distributions inside a NiO capacitor with a conducting filament. (a) A schematic picture showing the sample geometry for Au(30 nm)/Pt(10 nm)/NiO(60 nm)/Pt(10 or 200 nm)/TiO$_x$(10 or 20 nm)/SiO$_2$(1000 or 300 nm) capacitors. A single conducting filament is assumed with dimensions of 30 × 30 × 60 nm located at the center of the capacitor. The Joule heat is assumed to be generated by current in the filament due to external voltage. (b) Time-dependent temperature rise at the center of the filament after the external voltage is applied. Note that temperature become saturated within about 1 μs. (c) and (d) Temperature contours for the cases with $t_{BE}$ = 200 nm and 10 nm, respectively. Note that temperature for the $t_{BE}$ = 10 nm case is about 200 K greater than that for the $t_{BE}$ = 200 nm case.

Fig. 3 (color online). Vector representations of the heat flow, i.e., $-k\nabla T$, for cases with (a) $t_{BE}$ = 200 nm and (b) $t_{BE}$ = 10 nm. Note that the heat flow vectors are more concentrated for the thinner bottom electrode, indicating a less efficient heat flow.

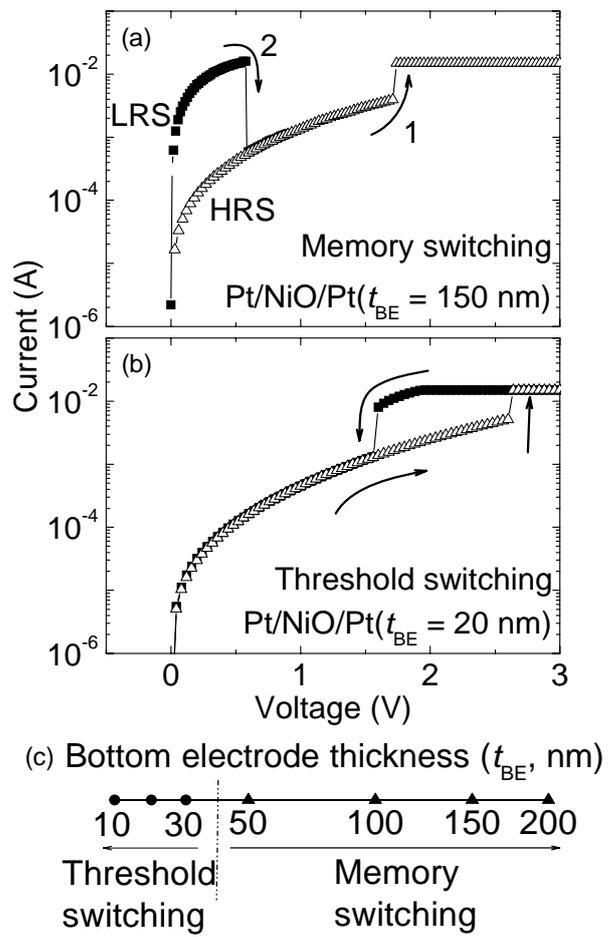

Figure 1

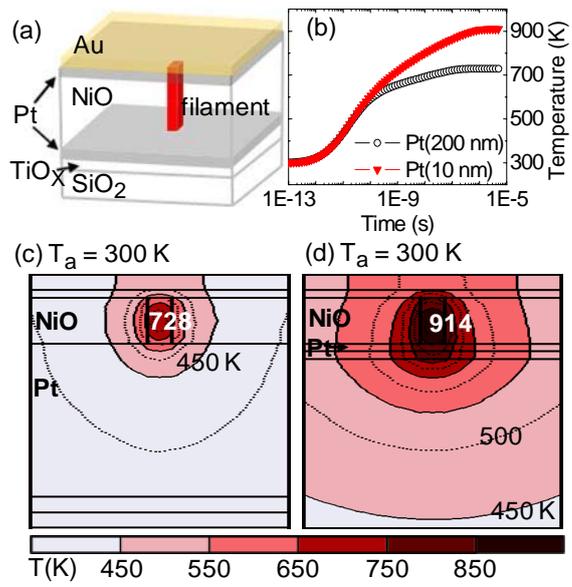

Figure 2

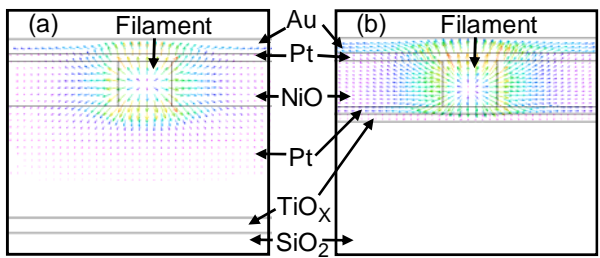

Figure 3